\documentclass{ws-mpla}

\usepackage{latexsym,amssymb,cite}

\pdfoutput=1

\newcommand{\cref}[1]{Chapter~\ref{#1}}
\newcommand{\bcenter}{\begin{center}}
\newcommand{\ecenter}{\end{center}}
\newcommand{\bean}{\begin{eqnarray*}}
\newcommand{\eean}{\end{eqnarray*}}
\newcommand{\ba}{\begin{array}}
\newcommand{\ea}{\end{array}}
\newcommand{\ben}{\begin{enumerate}}
\newcommand{\een}{\end{enumerate}}
\newcommand{\bd}{\begin{description}}
\newcommand{\ed}{\end{description}}

\def\IZ{\mathbb{Z}}

\def\IP{\mathbb{P}}

\def\im{{\mathop {\rm im}}}

\def\cA{{\mathcal A}}
\def\cN{{\mathcal N}}

\def\cO{{\mathcal O}}

\def\to{\rightarrow}

\def\theequation{\thesection.\arabic{equation}}

\newcommand{\mat}[1]{\left[ \begin{matrix} #1 \end{matrix} \right]}

\begin{document}

\markboth{Yang-Hui He}{An Algorithmic Approach to Heterotic String Phenomenology}

\catchline{}{}{}{}{}

\title{An Algorithmic Approach to String Phenomenology}

\author{Yang-Hui He}

\address{
Rudolf Peierls Centre for Theoretical Physics, \\
Oxford University, 1 Keble Road, Oxford, OX1 3NP;\\
Merton College, Oxford, OX1 4JD;\\
Department of Mathematics, City University, London,\\
Northampton Square, London EC1V 0HB, U.K.
}

\maketitle


%

\begin{abstract}
We review the recent programme undertaken\footnote{Based on various talks by the author, given lately in Providence, Tucson, London, Potsdam and Fort Lauderdale, and on joint efforts by the Oxford Group for algebro-geometric methods in string phenomenology, to which the author is most grateful for countless enlightening and jovial discussions and collaborations.} to construct, systematically and algorithmically, large classes of heterotic vacua, as well as the search for the MSSM therein.
Specifically, we outline the monad construction of vector bundles over complete intersection Calabi-Yau threefolds, their classification, stability, equivariant cohomology and subsequent relevance to string phenomenology.
It is hoped that this top-down algorithmic approach will isolate special corners in the heterotic landscape.
\end{abstract}

%


%
%
%
\section{Triadophilia: A Twenty-Year-Old Challenge}
Ever since the realization \cite{CHSW}, shortly after the discovery of the heterotic string \cite{het}, that it provides a promising venue wherein the Standard Model may be embedded, string phenomenology was born.
Indeed, the $E_8$ gauge group of the heterotic string encompasses a natural gauge unification of the Standard Model $SU(3) \times SU(2) \times U(1)$ group.
A mathematically succinct and physically appealing approach was thereby engendered, where the $E_8$ string was compactified on a Calabi-Yau threefold $X$.
The tangent bundle $T_X$ of $X$ admits, by the Calabi-Yau nature, an $SU(3)$ connection and $E_8$ is thus broken to an $E_6$ GUT group, the commutant of $SU(3)$ therein. This was the initial setup and the $E_6$ particle content is conveniently computed by the cohomology groups taking value in $T_X$, which, by standard Hodge decomposition of K\"ahler manifolds \cite{Hart,GH}, are simply the Hodge numbers of $X$:
\begin{equation}
n_{27} = h^1(X,T_X) = h^{2,1}_{\bar{\partial}}(X) \ , \quad
n_{\overline{27}} = h^1(X,T_X^*) = h^{1,1}_{\bar{\partial}}(X) \ .
\end{equation}
Now, the 27 of an $E_6$ GUT theory contains the entirety of the Standard Model fermions, hence the net number of generations is simply (the absolute value of) the difference between the above two terms, which, by rudimentary topology, is half the Euler number of $X$.

Our desire, is that there be three net generations, in accord with experimental observations.
Therefore, one of the first questions in string phenomenology was whether there exists a Calabi-Yau threefold (CY3) with Euler number $\pm 6$:
\begin{equation}\label{Euler6}
|\chi(X)| = 2|h^{1,1}(X) - h^{2,1}(X)| \stackrel{?}{=} 6 \ . 
\end{equation}
This, was the genesis of {\it Triadophilia} \cite{Candelas:2007ac} (Gk., love of three-ness).

The beautiful work of Candelas et al.~attempted to address this problem \cite{cicy,cicy2,cicy3,cicy4,cicy5,hubsch} in the first decade after the incipience of heterotic model-building. This was a pre-cursor to an algorithmic approach to string phenomenology and a large data-set, of 7890 manifolds, was constructed.
These are the so-called {\bf CICY manifolds}, or complete intersection Calabi-Yau threefolds embedded as $K$ homogeneous polynomials in $\IP^{n_1} \times \ldots \times \IP^{n_m}$. Here, complete intersection means that the dimension of the ambient space exceeds the number $K$ of defining equations by precisely 3, i.e., $K=\sum\limits_{r=1}^m n_r-3$. Moreover, the Calabi-Yau condition of vanishing first Chern class of $T_X$ translates to $\sum\limits_{j=1}^K q^{r}_{j} = n_r + 1 \ \forall \; r=1, \ldots, m$.
Subsequently, each manifold can be written as an $m \times K$ configuration matrix (to which we may sometimes adjoin the first column, designating the ambient product of projective spaces, for redundant clarity):
\begin{equation}
X = 
\left[\ba{c|cccc}
  \IP^{n_1} & q_{1}^{1} & q_{2}^{1} & \ldots & q_{K}^{1} \\
  \IP^{n_2} & q_{1}^{2} & q_{2}^{2} & \ldots & q_{K}^{2} \\
  \vdots & \vdots & \vdots & \ddots & \vdots \\
  \IP^{n_m} & q_{1}^{m} & q_{2}^{m} & \ldots & q_{K}^{m} \\
  \ea\right]_{m \times K \ ,}
\quad
\begin{array}{l}
K = \sum\limits_{r=1}^m n_r-3 \ , \\
\sum\limits_{j=1}^K q^{r}_{j} = n_r + 1 \ , \ \forall \; r=1, \ldots, m \ .
\end{array}
\end{equation}

The authors of \cite{cicy,cicy2,cicy3,cicy4,cicy5} proved that there can be only a finite number of such matrices and spent a number of years, in the nascent days of computer power, to establish a complete classification.
The most famous example is, of course, the manifold $[4|5]_{-200}^{1,101}$ (or simply $[5]$), known commonly as the {\bf quintic}. We have marked the Hodge numbers as superscript and Euler number, subscript.

Unfortunately, none of the 8000 or so CICYs has Euler number $\pm 6$.
Nevertheless, it was soon realized by Tian and Yau that freely-acting quotients of these spaces could have the correct property and the manifold $M = \mat{1&3&0\\ 1&0&3\\} / \IZ_3$  has topological numbers $M_{-6}^{6,9}$.
It was thus not surprising that this manifold became central to string phenomenology in the early days \cite{earlypheno,earlypheno2,earlypheno3}.

Nowadays, $E_6$ GUT theories are less favoured than their $SU(5)$ or $SO(10)$ counterparts, which can then be broken to the (Minimally Supersymmetric) Standard Model (MSSM) by turning on discrete Wilson lines.
Furthermore, $M_{-6}^{6,9}$ has 6 entire generations of anti-families, another cumbersome feature.

Modern heterotic phenomenology focuses on $SU(5)$ or $SO(10)$ GUTs and their subsequent breaking to the MSSM.
Now, obtaining these gauge groups is relatively straight-forward: one simply recalls that their commutants in $E_8$ are, respectively, $SU(5)$ and $SU(4)$.
Therefore, endowing the CY3 $X$ with an $SU(n)$ bundle $V$ with $n=3,4,5$ will generalize the traditional setup where $V= T_X$ (now known as ``standard embedding'') to give these GUT theories (now known as ``general embedding'') \cite{newissue,GSW}.
The adjoint 248 of $E_8$ branches accordingly:
\[
\mbox{
\begin{tabular}{|l|l|}
\hline
$E_{8}\rightarrow G\times H$ & Residual Group Structure \\  \hline\hline
$\rm{SU}(3)\times E_{6}$ & $248\rightarrow (1,78)\oplus (3,27)\oplus (\overline{
3}
,\overline{27})\oplus (8,1)$ \\  \hline
$\rm{SU}(4)\times\rm{SO}(10)$ &$ 248\rightarrow (1,45)\oplus (4,16)\oplus (\overline{4
},\overline{16})\oplus (6,10)\oplus (15,1)$ \\  \hline
$\rm{SU}(5)\times\rm{SU}(5)$ & $248\rightarrow (1,24)\oplus (5,\overline{10})
\oplus (
\overline{5},10)\oplus (10,5)\oplus (\overline{10},\overline{5})\oplus (24,1)$
 \\ \hline
\end{tabular}
}
\]
The associated particle content is likewise captured by the various vector-bundle-valued cohomology groups:
\[\mbox{
\begin{tabular}{|l|l|} \hline
Decomposition & Particle Content \\ \hline
$\rm{SU}(3)\times E_{6}$ & $n_{27}=h^{1}(V), n_{\overline{27}}=h^{1}(V^{\ast
})=h^{2}(V), n_{1}=h^{1}(V\otimes V^{\ast })$ \\ \hline 
$\rm{SU}(4)\times \rm{SO}(10)$ &$ n_{16}=h^{1}(V), n_{\overline{16}}=h^{2}(V), 
n_{10}=h^{1}(\wedge ^{2}V), n_{1}=h^{1}(V\otimes V^{\ast })$ \\ \hline 
$\rm{SU}(5)\times\rm{SU}(5)$ & $n_{10}=h^{1}(V^*), n_{\overline{10}}=h^{1}(V), 
n_{5}=h^{1}(\wedge^{2}V), n_{\overline{5}}=h^{1}(\wedge ^{2}V^{\ast })$\\
&$n_{1}=h^{1}(V\otimes V^{\ast })$ \\ \hline
\end{tabular}
}\]
Next, breaking the GUT to the MSSM gauge group is also elementary (what is difficult, as we shall see, is obtaining the precise spectrum and couplings):
one enriches the above structure of the pair $(X,V)$ with a discrete Wilson line $W$ such that the commutant of $W$ in $SU(5)$ or $SO(10)$ is $SU(3) \times SU(2) \times U(1)$ (with a possibility of an extra $U(1)$ factor).
On a Calabi-Yau manifold with non-trivial (discrete) fundamental group $W = \pi_1(X)$, one can turn on a $W$-Wilson line and compute the $W$-equivariant cohomology groups for $V$, in conjunction with the action of the Wilson line, to obtain the final particle spectrum.
This has been developed in a programme led by Ovrut, Donagi, Pantev et al.over the past decade: cf.~\cite{Lukas:1998yy,Donagi:2003tb,Donagi:2004ia,Donagi:2004qk} (q.v.~brief review in \cite{He:2005hz}) and \cite{Donagi:2004su,Donagi:2004ub,Braun:2005ux,Braun:2005bw,Braun:2005zv}.
This search for the MSSM, wherein the particle content is encaptured by bundle cohomology, the moduli, bundle endomorphisms, and Yukawa couplings, trilinear compositions amongst the cohomology groups, has been a substantial challenge and a healthy dialogue for collaborative efforts between physicists and mathematicians.

%
%
%
\section{A Special Corner}
A preliminary success of the aforementioned endeavour culminated in 2005 where, sifting through some number of suitable vector bundles over so-called elliptically-fibered CY3s (another large class of manifolds), two complementary heterotic MSSM models with no exotic particles, no anti-generations, exactly one pair of Higgs and reasonable Yukawa couplings were found after years of search.
One \cite{Braun:2005nv,Braun:2006me,Braun:2006ae} was based on an $SU(4)$ bundle coupled with a $\IZ_3 \times \IZ_3$ Wilson-line, giving rise to, via an $SO(10)$ GUT, the MSSM with an $U(1)_{B-L}$, the other \cite{Bouchard:2005ag,Bouchard:2006dn}, on an $SU(5)$ bundle coupled with a $\IZ_2$ Wilson-line, giving the MSSM via an $SU(5)$ GUT.
Both models are based (respectively by equivariant $\IZ_3 \times \IZ_3$ and $\IZ_2$ actions) on a parent self-mirror $X^{19,19}_0$ manifold, which is a $T^2$-fibration over a nineth del Pezzo surface, which is itself an elliptic fibration over $\IP^1$.

Remarkably, this $X^{19,19}_0$ covering manifold, from which both quotients descend, resides in the intersection between the database of elliptically fibered threefolds and CICYs and it was an astute observation, initially made by Candelas, that it is, in fact, intimately related to the cover of the Tian-Yau manifold:
\begin{equation}
X_0^{19,\,19} = \mat{1&1\\3&0\\0&3} \ , \quad
X_{-18}^{14,23} = \mat{1&3&0\\ 1&0&3\\} \ .
\end{equation}
Why indeed should the two most favoured manifolds from two completely different data-sets, spanning over a decade of search, be connected by so simple a transposition of configuration matrices?

The {\it observatio curiosa} prompted the investigations in \cite{Candelas:2007ac} wherein the two manifolds, together with the bi-cubic CICY, $X_{-162}^{2,83} = \mat{3\\3}$, as well as their various quotients were shown to be conifold transition of each other.
The bundles thereon, were proposed to be linked by so-called {\bf transgressions}, where a bundle on one is taken, being trivial on the conifold blowup $\IP^1$ cycles, to another bundle on the other after the conifold transition.
Indeed, all CICYs are related by conifold transitions and it is a conjecture known as {\bf Reid's fantasy} that all CY3s are related by blow-up/down transitions generalizing the conifold; it is thus natural to speculate that all (stable) vector bundles on all CY3s should transgress -- perhaps the plethora of heterotic vacua are all inter-laced after all!

To have a glimpse of this multitude, a plot was made of all the known CY3s to date in \cite{Candelas:2007ac}, and in the spirit of \cite{cicy,cicy2,cicy3,cicy4,cicy5} as well as the pioneering work in mirror symmetry by Candelas et al., the ordinate and abscissa are taken, respectively, as the sum and the difference (Euler number) of the two Hodge numbers.
\begin{figure}[t]
\begin{center}
(a)
\includegraphics[scale=0.7]{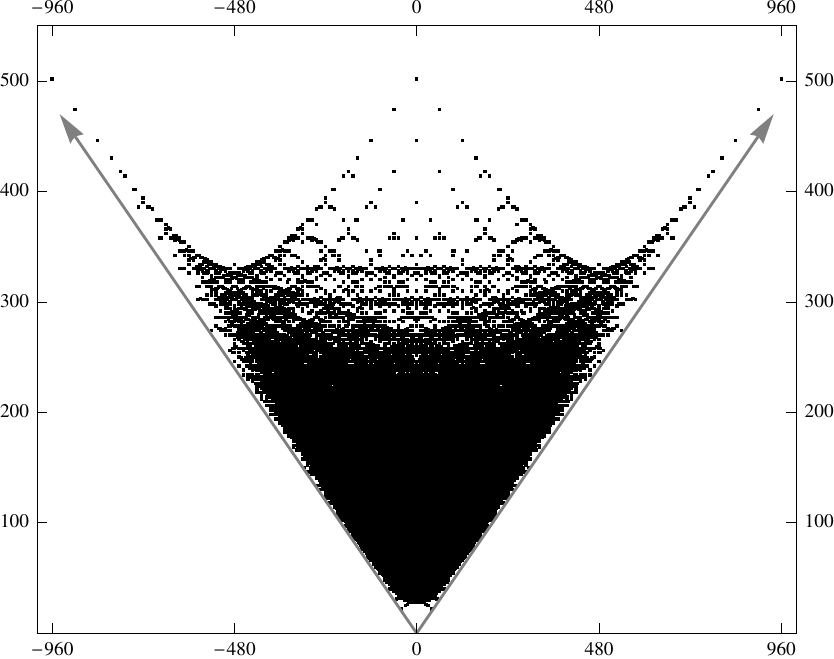}
(b)
\includegraphics[scale=0.6]{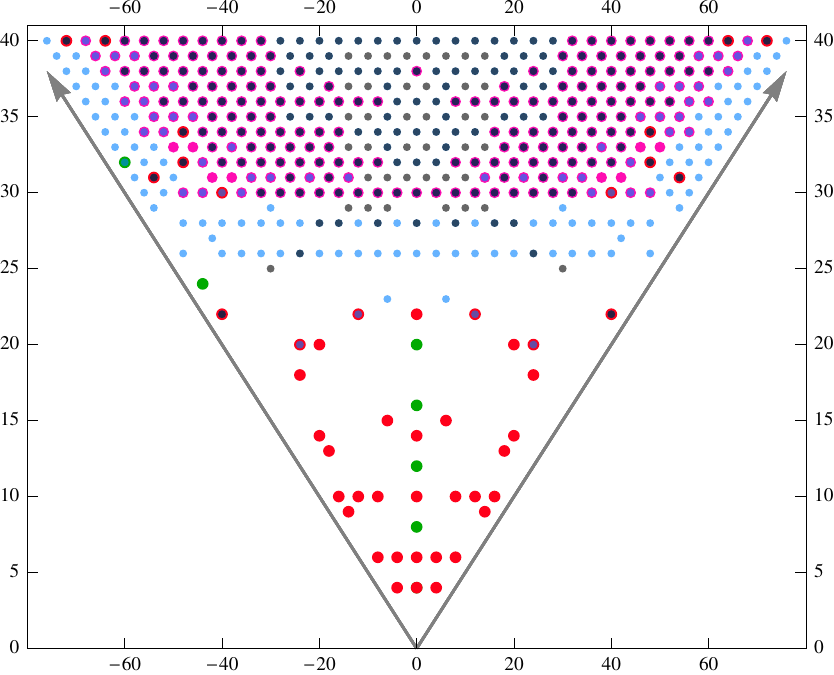}
\caption{{\sf (a) Plotting $\chi = 2(h^{1,1}-h^{1,2})$ (horizontal) versus $h^{1,1} + h^{1,2}$ (vertical) of all the some $5 \times 10^8$ known Calabi-Yau threefolds; (b) magnification of the bottom corner in (a).}}
\label{f:CYplot}
\end{center}
\end{figure}
This is presented in Part (a) of Figure \ref{f:CYplot}.
There are about 500 million known Calabi-Yau manifolds, including the CICYs, the elliptically fibered, isolated examples, etc.; the largest family by far, however, is the impressive list compiled over the 1990's, of smooth hypersurfaces in toric four-folds, performed in the lovely work of Kreuzer-Skarke \cite{Kreuzer:2000xy,Avram:1997rs}, consisting of 473,800,776 inequivalent CY3s, with 30,108 distinct pairs of Hodge numbers contributing to Figure \ref{f:CYplot} (a).

There are certainly many intriguing observations to be made on the plot, such as whether there can exist a CY3 whose Euler number exceeds 960 in magnitude\footnote{The author has bet a fine bottle of port with Dr.~Andrew Dancer of Jesus College, Oxford, a wager recorded in the antiquate Betting Book of the said College, the first pages of the present Volume consisting of entries concerning the forces of Napoleon, that there does {\it not} exist such a Calabi-Yau threefold.}.
More immediate to our concern, however, is that the centre of the plot has multiple occupation, that is, each point represents up to hundred of thousands of CY3s.
Nevertheless, as two decades of construction have taught us, in conjunction with subsequent work in \cite{Candelas:2008wb,Braun:2009qy}, that the bottom corner of the plot is rather sparse.
Zooming into this {\bf special corner}, presented in part (b) of the Figure, shows rather pretty patterns.
Importantly, these low-Hodge-number manifolds comprises of the few ``good'' examples on which bundles begetting exact MSSM spectrum live and mutually transgress.
Is this corner an {\it oasis} in the seeming supererogation of string vacua, wherein a world like ours could blossom, while the remaining plenitude are no more than a swap-land \cite{Vafa:2005ui,Gmeiner:2005vz}?

\section{Algorithmic Scan of Vacua}
Short of a {\it selection principle}, an immediate method of approach to the special corner is not obvious, instead, a synthetic rather than analytic perspective may prove to be conducive.
This has been undertaken over the past few years \cite{Anderson:2007nc,Anderson:2008uw,Anderson:2008ex,Anderson:2009ge,He:2009wi,Anderson:2009nt,Anderson:2009sw,Gabella:2008id,Anderson:2009mh}, where large classes of bundles are constructed over large data-sets of CY3s and then those with MSSM properties, carefully selected.
Such a philosophy of algorithmic scan is indeed facilitated by the rapid advances in computational algebraic geometry as well as its implementation on ever-faster machines \cite{sing,m2}; the cross-pollination of this subject with theoretical and mathematical physics has also recently been a healthy endeavour \cite{Gray:2006jb,Gray:2008zs,He:1900zza}.

\subsection{Physical and Mathematical Constraints}\label{s:cons}
We therefore wish to explicitly construct a large number of special unitary vector bundles $V$ on CY3s; quite a few physical constraints can instantly be imposed.
\paragraph{Supersymmetry: }
We are aiming for the MSSM, and hence require $\cN = 1$ SUSY in the low-energy 4-dimensional theory, which in turn implies that $V$ admits a holomorphic connection $F$ satisfying the Hermitian-Yang-Mills (HYM) equations: $ F_{ab} = F_{\bar{a}\bar{b}} = g^{a\bar{b}}F_{a\bar{b}} = 0$, a generalization of Ricci-flatness for Calabi-Yau manifolds.
Solving these PDEs explicitly are currently impossible.
Luckily, we are saved by the Donaldson-Uhlenbeck-Yau theorem \cite{D,UY}, which states that on each (poly-)stable holomorphic vector bundle, there exits a unique HYM connection.

Let us not delve into the details of stability, but only emphasize that this can be construed as a purely algebraic condition, without recourse to hard analysis.
Some immediate and calculationally important consequences are that the zeroth and highest cohomologies of a stable bundle $V$ vanish: $H^0(X,V) = H^3(X,V) = 0$; this, coupled with {\bf Serre duality} for any bundle $V$ on a CY3, that $H^p(X,V) \simeq H^{3-p}(X,V^*)^*$, imply the following \cite{Hart,GH}:
\begin{equation}\label{stab}
H^0(X,V) = H^0(X,V^*) = H^3(X,V) = H^3(X,V^*) = 0 \ .
\end{equation}
\paragraph{Net Generations: }
The vanishing conditions \eqref{stab}, together with the Atiyah-Singer index theorem on $X$ which generalizes the statement for the Euler number, yield that
\begin{equation}
\mbox{index}(\slash \!\!\!\!\nabla_X) = \sum\limits_{i=0}^3 (-1)^i
h^i(X,V) = \int_X \text{ch}(V) \text{td}(X) = \frac12 \int_X c_3(V) \ .
\end{equation}
Consequently, this gives us an expression for the net number of generations of particles, analogous to \eqref{Euler6}:
\begin{equation}
N_{gens} = -h^1(X,V) + h^1(X,V^*) = \frac12 \int_X c_3(V) \ .
\end{equation}
We will require that this number be a multiple of three, say $3k$, where $k$ is the order of a possible freely acting group $G$ on $X$, so that upon descending to the quotient manifold $X/G$, there would be precisely 3 generations.
Indeed, in order that $G$ be a free action, $k$ must necessarily (but not sufficiently) divide the Euler number $\chi(X)$.
\paragraph{Anomaly Cancellation: }
To ensure Green-Schwarz anomaly cancelation \cite{GS}, it is standard to set $\int_X R \wedge R - F \wedge F = 0$, where $R$ is the Ricci form on $X$, that is, $c_2(X) = c_2(V)$.
However, one could allow M5-branes in the bulk, in a heterotic M-theory Ho\v{r}ava-Witten setup \cite{HW,Donagi:1999ez}, which could wrap effective holomorphic 2-cycles (i.e., actual curves).
Hence, one could allow that the difference $c_2(X) - c_2(V)$ correspond to an effective class.

Finally, since we are dealing with SU-bundles, $c_1(V) = 0$.
Thus, in sum, we have constraints on all the three Chern classes of $V$.

\subsection{Monad Bundles: A Large Class}
Our strategy is clear.
We will attempt to construct a large class of bundles satisfying the above constraints over a substantial data-set of CY3s.
The most systematic method is the so-called {\bf monad construction} over projective varieties \cite{monadbook}, developed by algebraic geometers in the later half of the last century. In fact, all vector bundles over projective spaces can be so obtained, which is perhaps why such a Leibnizian name, with regard to the universality implied, was originally chosen.
It is thus not surprising that these bundles have been used sporadically but fruitfully in a physical context since the early days of string model building \cite{Distler:1987ee,Kachru:1995em,Blumenhagen:1996vu,Blumenhagen:1997vt,Douglas:2004yv,maria}.

The building block to a monad is a {\bf line bundle}, the simplest vector bundle on any manifold. On CICYs, a line bundle is easy to describe and so let us begin with this historical data-set \cite{Anderson:2007nc,Anderson:2008uw,Anderson:2008ex,Anderson:2009ge,Anderson:2009nt,Anderson:2009sw,Anderson:2009mh}.
Denote  $\mathcal{O}_{\IP^n}(k)$, as is standard, the $k$-th power of the hyperplane bundle $\mathcal{O}(1)$ over $\IP^n$, then its first chern class is $c_1(\mathcal{O}(k)) = k J$ with $J$ being the K\"ahler class of $\IP^n$.
Subsequently, one could restrict this to any projective variety $X$ to obtain $\mathcal{O}_{X}(k)$ with the proviso that $X$ also has only a single K\"ahler class, as descended from the ambient $\IP^n$.
Such cases of when $h^{1,1}$ of the ambient and variety itself are both unity is called {\bf cyclic}.

In general, our ambient space $\cA$ for CICYs are products of $m$ projective spaces in which $K$ homogeneous polynomials define $X$. 
We shall call the cases where $h^{1,1}(X) = h^{1,1}(\cA) = m$ as {\bf favourable}; here the K\"ahler classes descend completely from $\cA$ to $X$.
In this case we can write line bundles over $\cA = \mathbb{P}^{n_1 }\times \mathbb{P}^{n_2} \times \ldots \times \mathbb{P}^{n_m}$ as $\cO_{\cA}(k_1,k_2,...,k_m)$ with corresponding restriction to the CICY, $X$.
Finally, we have the Kodaira vanishing theorem \cite{Hart,GH} which states that for a positive line bundle $P$ on a Calabi-Yau manifold $X$ , 
$H^q(X,P)=0$ for all $q>0$.
For CICYs, a positive line bundle is one for which each $k_i$ entry above is positive.
This vanishing will be of significant aid to us in computing the spectrum and couplings later.
Therefore, our starting point will be monads constructed from positive line bundles over favourable CICYs.

Thus prepared, we can define a monad as the bundle $V$ which resides in a short exact sequence (i.e., a free resolution of length 2):
\begin{equation}
0 \to V \stackrel{f}{\longrightarrow} B \stackrel{g}{\longrightarrow}C \to 0 \ ;
\mbox{ with }
\qquad
{\scriptsize
B=\bigoplus\limits_{i=1}^{r_B} \cO(b_{r}^{i})\; ,\qquad
C=\bigoplus\limits_{j=1}^{r_C} \cO(c_{r}^j) \ .
}
\end{equation}
Here, short exactness implies that $V = \im(f) \simeq ker(g)$ and that $\mbox{rk}(V)=\mbox{rk}(B)-\mbox{rk}(C)$.
The map $g$ is explicitly a matrix of polynomials; e.g., on $\mathbb{P}^n$ the $ij$-th entry is a homogeneous polynomial of degree $c_i - b_j$.
Moreover, our positivity requirement implies that all integers $b^i_r, c^j_r$ be strictly positive.

In summary, the physical constraints in \S\ref{s:cons} manifest themselves as a list of combinatorial conditions on the integers $b^i_r, c^j_r$:
\begin{enumerate}
\item Bundle-ness: $b^{i}_r \leq c^{j}_r$ for all $i,j$ and the map $g$ can be taken to be {\it generic} so long as exactness of the sequence is ensured;
\item SU-Bundle: $c_1(V)=0  \Leftrightarrow \sum \limits_{i=1}^{r_B} b^{r}_i -\sum \limits_{j=1}^{r_C} c^{r}_j = 0$;
\item Anomaly cancellation: $c_2(X)- c_2(V) = c_2(X)-\frac12 (\sum\limits_{i=1}^{r_B} {b^{i}_s}{b^{i}_t} - \sum\limits_{i=1}^{r_C} {c^{j}_s}c^{t}_i)J^{s}J^{t} \geq 0$;
\item Three Generations: $c_3(V) =\frac{1}{3} (\sum\limits_{i=1}^{r_B}{{b_r}^i}{{b_s}^i}{{b_t}^i} -  \sum\limits_{j=1}^{r_C}{{c_r}^j}{{c_s}^j}{{c_t}^j}) {J^r}{J^s}{J^t} = 3k$, with $k$ a divisor of $\chi(X)$ \ .
\end{enumerate}

Upon finding such monads, we shall then prove stability and compute the particle spectrum and Yukawa couplings.
These are extremely technical and difficult problems and have been a major hindrance to existing literature if one were to calculate them for large classes on the order of thousands or more.
A break-through in the method of attack of the programme in \cite{Anderson:2007nc,Anderson:2008uw,Anderson:2008ex,Anderson:2009ge,He:2009wi,Anderson:2009nt,Anderson:2009sw,Gabella:2008id,Anderson:2009mh} is the extensive use of the recent rapid advances in computer algebra, especially in algorithmic algebraic geometry \cite{sing,m2}.
We shall not dwell too much in this review on these technicalities and shall only mention that for stability, at least for cyclic manifolds, a so-called Hoppe's criterion reduces the problem to computing cohomology groups.
To arrive at the various cohomologies, the standard approach here is to use spectral sequence induced from the Koszul resolution of $V$, whereby simplifying the problem to known cohomologies (via Bott-Borel-Weil and K\"unneth \cite{Hart,GH}) on the ambient product of projective spaces.

\begin{figure}[h!!!]
(a) \includegraphics[trim=0mm 0mm 100mm 200mm, clip, width=2.2in]{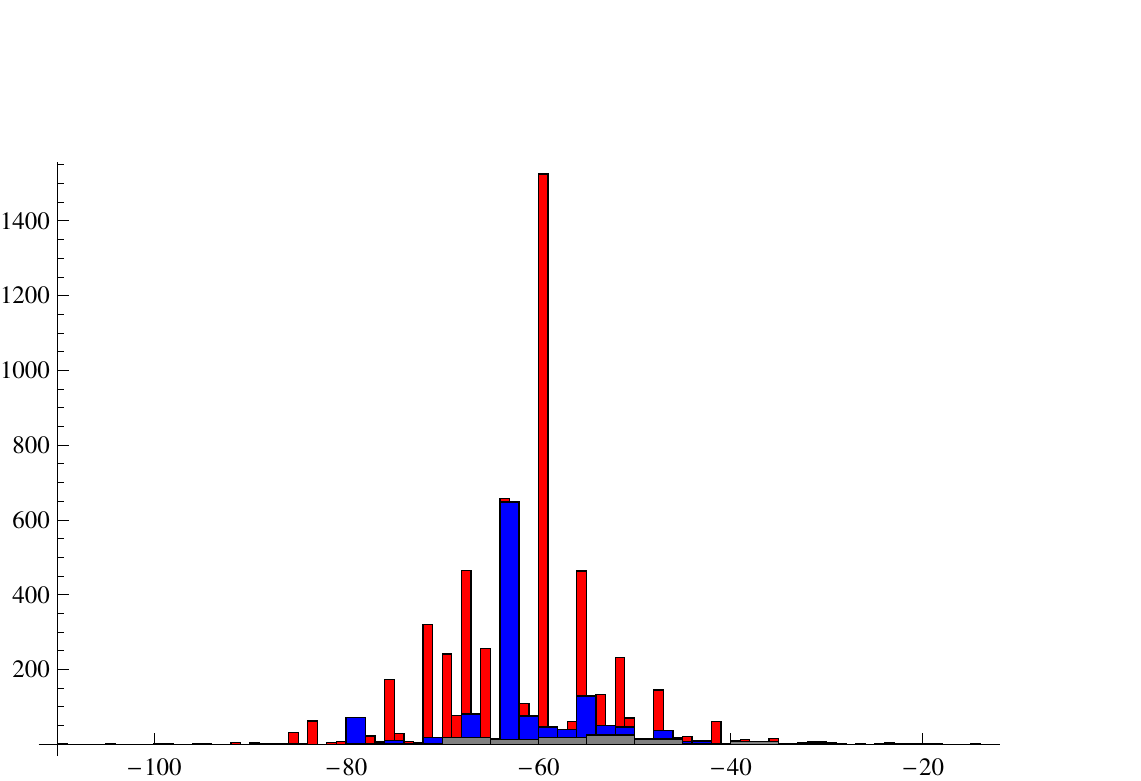}
(b) \includegraphics[trim=0mm 0mm 100mm 200mm, clip, width=2.2in]{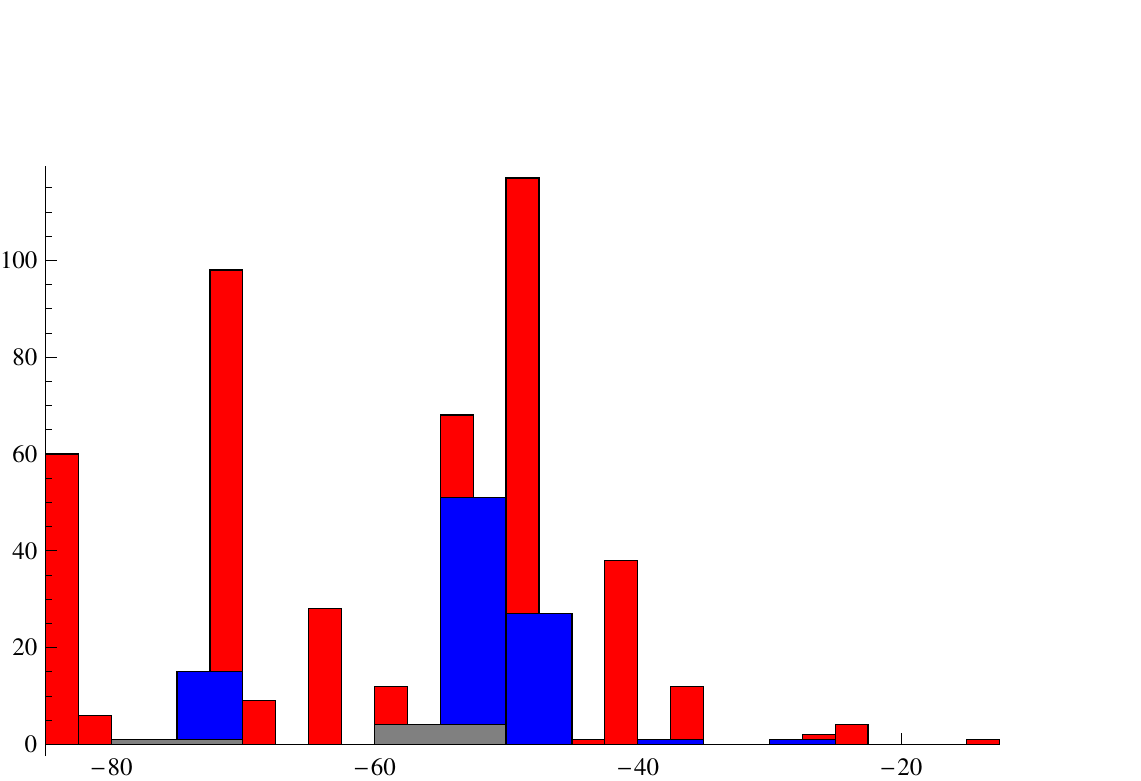} 
\caption{{\sf (a) Histogram for the index,
${\rm ind}(V)$, of the 7118 positive monads
found over 36 favourable CICYs: the horizontal axis is ${\rm ind}(V)$ and the
vertical, the number of bundles;
(b) the same data set,
but only taking into account monads with ${\rm ind}(V) = 3k$ for some positive 
integer $k$, such that $k$ divides the Euler number of the corresponding CICY.
}}
\label{f:pos}
\end{figure}

It was shown in \cite{Anderson:2008uw} that positive bundles satisfying the above list of constraints is finite in number, amounting to about 7000 on only a small number of CICYs.
We present the statistics thereof below and the histograms of $c_3(V)$, i.e., the number of generations, in Figure~\ref{f:pos}:
\begin{tabular}{|c|c|c|c|c|}  \hline
  & Bundles & ${\rm ind}(V) = 3k$ & \begin{tabular}{l} ${\rm ind}(V) = 3k$ \\ an
d $k$ divides $\chi(X)$ \end{tabular} 
  & \begin{tabular}{l} ${\rm ind}(V) = 3k$ \\ $|{\rm ind}(V)|<40$ \\
  and $k$ divides $\chi(X)$ \end{tabular} \\ \hline \hline
rank 3 & 5680 & 3091 & 458 & 19\\
rank 4 & 1334 & 207 & 96 & 2 \\
rank 5 & 104 & 52 & 5 & 0 \\ \hline
Total & 7118 & 3350 & 559 & 21\\ 
\hline
\end{tabular}
\subsection{A Needle in A Haystack}
Emboldened by our ability to classify and further intrigued by the rarity of occurrence of anything akin to the MSSM, we enlarged our data-set to include the so-called semi-positives; these are monads with possible zeros in the entries.
Currently, short of strong isomorphism theorems we do not have a finiteness result even though the numbers are expected to be so; preliminary scans have produced a substantially larger set.
On these bundles an initial scan was performed and single one was so far found \cite{Anderson:2009mh} to have the exact MSSM spectrum\footnote{The author is most obliged to Mlle.~N.~Davis, {\it Zuleika Mertonensis}, for the charming diversions afforded unto him during this work.}.
Interestingly, the base manifold turned out to be the bi-cubic, $X=$  {\tiny $\left[\begin{array}[c]{c}\mathbb{P}^2\\\mathbb{P}^2\end{array}\left|\begin{array}[c]{ccc}3 \\3 \end{array}\right.  \right]^{2,83}_{-162}$}, again one of the guises of the Tian-Yau manifold!

Specifically, this new heterotic standard model, to be added to the other two within the special corner, has no anti-generations, no exotic particles, one pair of MSSM Higgs and an $U(1)_{B-L}$.
It is obtained from an $SU(4)$ monad bundle
\begin{equation}\label{Vmssm}
0 \to V \to \cO_X(1,0)^{\oplus 3} \oplus \cO_X(0,1)^{\oplus 3} \stackrel{f}{\rightarrow} \cO_X(1,1) \oplus \cO_X(2,2) \to 0 \ ,
\end{equation}
giving first an $SO(10)$ GUT, which is subsequently broken by a $\IZ_3 \times \IZ_3$ Wilson line to give the desired $SU(3) \times SU(2) \times U(1) \times U(1)_{B-L}$ group.
In order to turn on this Wilson line we ensured that there exists a freely acting $\IZ_3 \times \IZ_3$-group, which was found in \cite{Candelas:2007ac}, giving a quotient manifold $X^{2,11}_{-18}$.
To guarantee that the bundle $V$ also descends to the quotient, it was further shown that \eqref{Vmssm} admits an equivariant structure with respect to the group action, and hence the quotient is also a bona fide bundle.

\section{Outlook}
Armed with the advances in contemporary algebraic geometry, encouraged by the possibility of extensively utilizing computer algebra coupled with physical intuition and insight in order to sift through vast data-sets, and spurred by the observation and speculation of transgressions within a special corner in the space of stringy vacua, our spirits are enheartened by optimism.
For the first time, the prospects of an systematic and thorough expedition of the heterotic landscape, a venture dreamt of for two decades, seem within our grasp. We have explored the oldest geography of the CICYs, and already found a rare occurrence of the MSSM; meanwhile, reconnaissance into the further territory of the elliptic manifolds and most importantly the toric hypersurfaces are well under way \cite{Gabella:2008id,He:2009wi}.
The glimpse into a barren expanse of a {\it terra incognita}, harbouring perhaps a rare treasure cove bejewelled by universes close to our own, enlivens our souls. Onwards we must march.

\section*{Acknowledgements}
{\it Ad Catharinae Sanctae Alexandriae et Ad Majorem Dei Gloriam...\\}
The author would like to lend this opportunity to gratefully acknowledge his Oxford friends L.~Anderson, P.~Candelas, X.~de la Ossa, J.~Gray, S.-J.~Lee, A.~Lukas, B.~Szendr\"oi, as well as his old Philadelphia confreres V.~Braun, R.~Donagi, B.~Ovrut, and T.~Pantev for many years of fruitful and enjoyable collaborations on this subject. 
He is most indebted to the gracious patronage of Merton College, Oxford, whose Fellowship bestowed upon him is held dearly to his heart, and he carouses to the health of the parting Warden, Professor Dame Jessica Rawson as he salutes the new, Professor Sir Martin Taylor. {\it Stet fortuna Domus!}


\end{document}